\documentclass[10pt,a4paper,twoside]{article}

\usepackage{indentfirst}                        % Required!
\usepackage{graphicx}                           % Required if figures
\usepackage{amsgen,amsfonts,amssymb,amsbsy}     % Enlarge math symbols set

\usepackage{epsf}
\usepackage{amsmath}
\usepackage{amssymb}
\usepackage{amsthm}

\newenvironment{resum}{\begin{quote}\small}{\end{quote}}

\newcommand{\bfsf}[1]{\textsf{\textbf{#1}}}

\begin{document}

\thispagestyle{plain}           % Remove headings in first page

\begin{center}

%  Title; use linebreaks with "\\" if necessary:

{\LARGE\bfsf{Computer Algebra Applications\\ 
             for Numerical Relativity}}

\bigskip

%  Author list:

\textbf{Sascha Husa}$^{1,2}$ and \textbf{Christiane Lechner}$^1$

%  Affiliation. State only name and Institution of authors:

$^1$\textsl{Max-Planck-Institut f\"ur Gravitationsphysik, Albert-Einstein-Institut, Germany} \\
$^2$\textsl{Departament de F\'{\i}sica, Universitat de les Illes Balears, Spain} \\

\end{center}

\medskip

%  Place abstract here:

\begin{resum}
We discuss the application of computer algebra to problems commonly arising in
numerical relativity, such as the derivation of 3+1-splits, manipulation of
evolution equations and automatic code generation. Particular emphasis is put
on working with abstract index tensor quantities as much as possible.
\end{resum}

\bigskip

% end of macro definitions
%
%
%
%\section{Introduction}

Numerical relativists rarely talk about computer algebra (CA) in public, and
outsiders to the field even might get the impression that CA
and numerical relativity (NR) indeed represent two antipodal uses of
computers to study relativity theory. Quite the opposite is true. 
In this article we will try to outline some of the possible applications 
of computer algebra for NR, and very briefly
present some of our recent work resulting in the creation of a suite of
{\em Mathematica} scripts which we have found extremely useful and which
are available on request.
While one of our aims is to make practitioners of the field more aware of the
opportunities, another is to help outsiders better understand the
problems faced in NR. These are often analytical in nature and sometimes closer to
mainstream mathematical relativity than expected:
There is much more to NR than coding up, say, the ADM equations \cite{Luis}:
$$
{\cal L}_{n} h_{ab} = 2 \alpha K_{ab},
   \quad
{\cal L}_{n} K_{ab} = \alpha_{;ab} + \alpha\left(2 K_{cb} {K_{a}}^{c} 
- K K_{ab} - \alpha R_{ab}\right)
$$
-- or any other particular evolution system one fancies.
The perspective of numerical approximation raises many new questions
about the Einstein equations, such as what happens to the constraints
in a free evolution scheme.
Mathematical analysis and NR experience have shown that
the Einstein equations have to be brought into a form which is suitable for
numerical treatment. Considering evolution problems, note that obtaining
a well-posed problem is not sufficient. Well posedness does {\em not}
rule out exponential growth which may result from constraint violating modes
or a bad gauge. Curing such problems typically requires modifying the
equations, and the analysis and coding of different systems of equations.
NR thus provides perfect problems for CA, such as: (i) 3+1 or 2+2 decompositions,
(ii) modification of equations by adding
constraints, changing variables, etc., (iii) derivation and analysis of
associated systems like the constraint propagation system, (iv)
linearization around exact solutions or (v) the generation of numerical
code from systems of equations.

All of these tasks can in principle be accomplished by component
calculations, as can be carried out quite conveniently and efficiently
by computer algebra systems like GRTensorII \cite{grtensor}, which allows
to enter expressions in abstract index notation and yields
results in component form. This is often what one wants, but thinking about
deriving and analyzing evolution systems, it is clear that
apart from the fact that this may result in very large calculations
requiring significant time and memory, this method is unwieldy and
not very intuitive.
Rather one would like to keep an abstract index notation as long as possible,
and in particular get results in this notation.

%\section{Design and Implementation of a Suite of {\em Mathematica} Scripts}
%%%%%%%%%%%%%%%%%%%%%%%%%%%%%%%%%%%%%%%%%%%%%%%%%%%%%%%%%%%%%%%%%%%%%

The final aim of computer algebra calculations in numerical
relativity will usually be the generation of code, in C or Fortran say,
so let us consider a minimal task list for code generation.
From our pint of view, the problem of generating finite-differenced 
(or otherwise discretized equations) roughly splits into the following steps:
\begin{enumerate}
\item Write or find a CA system capable of abstract index
tensor calculus.
\item Write a package to facilitate 3+1 splits and other
calculations in a 3+1 context.
\item Derive a system from a 3+1 split or transform a given 3+1 system in some way. 
At the end of this process the
desired system is given in terms of its dependent variables,
their time derivatives and spatial ordinary derivatives.
\item Translate the tensor expressions into components.
\item Replace the ordinary derivatives by some standard language, e.g.
 {\tt D2[h13]}.
\item Create discretized expressions, e.g.\\
%\begin{center}
 {\tt D2[h13] $\rightarrow$ ( h13(i, j+1, k) - h13(i, j-1, k) ) / dy}.
\item
Wrap this up by code needed to create a full executable program
\end{enumerate}
For the task of coding a simple given system, such as ADM,
steps 1 -- 3 might be considered overkill -- it is easy to type in
the desired equations in abstract form by hand.
For more complicated first order systems or for deriving associated
systems like the constraint propagation system, linearizations or perturbation
formalisms, these techniques potentially save a lot of valuable time.

As far as code generation is concerned, we decided to
generate code as complete Cactus \cite{cactus}
thorns. This choice yields an open and
{\em documented} infrastructure,
parallelization, clean I/O methods and allows easy interfacing
with a growing community writing NR Cactus applications.
Modifications to interface with other systems with capabilities similar to
those of cactus or with home-brewed code  should be rather straightforward.

For the choice of CA-system we contemplated the use of {\em Maxima}, {\em Maple} and
{\em Mathematica}.
{\em Maxima} is a an open source version of {\em Macsyma},
and via the {\em itensor} package supports abstract index calculations.
However, to our knowledge {\em itensor} is currently not fully functional.
Provided development continues, {\em Maxima} could however become a very
interesting option.
{\em Maple} and {\em Mathematica} are both widely spread commercial
CA systems. Both provide support for component calculations, we find
the GRTensorII package for {\em Maple} particularly useful.
However, as opposed to {\em Mathematica}, we are not aware of any functional
abstract index tensor package for  {\em Maple}. We speculate that this is rooted
in {\em Mathematica}'s superior intrinsic support for
pattern matching. This seems quite essential for tensor manipulations,
e.g. $T_{ab}$ and $T_{cd}$ are not the same expression but
still are equivalent mathematical objects, which can easily
be identified with pattern matching techniques.
We have worked with two different abstract index packages for {\em Mathematica}:
the freely available {\em Ricci} and the commercial {\em MathTensor}.
%Conversion of tensor expressions between the two is possible, and we use this
%for a script to produce \LaTeX output from {\em MathTensor} expressions.
Despite our feeling that the overall design of {\em MathTensor} is less clean, and 
despite the fact that we have found several bugs and inconsistencies in
{\em MathTensor},
we still eventually selected it as the basis for our work. 
The main reasons were a
somewhat more extensive functionality, its support for both abstract and component
calculations, better checking for errors (such as inconsistencies with
indices), and it's simpler representation of tensors as plain functions of indices:
{\tt h[la, lb]} $\rightarrow$ $h_{ab}$,
{\tt CD[Metricg[la, lb], lc]} $\rightarrow$ $0$. This straightforward syntax
is less error-prone than {\em Ricci}'s corresponding {\tt h[L[a], L[b]]} or {\tt
h[L[a], L[b]] [L[c]]} for a covariant derivative.
Given {\em MathTensor}'s immense value, problems and significant cost, 
it would be very attractive to have available an open source alternative with similar
functionality.

The basic set of functions needed for 3+1 decompositions,
as well as some general tensor manipulation functionality, has been
implemented in a {\em Mathematica} notebook {\tt Decompose\_3+1\_Tools.nb}.
Our strategy to do 3+1 decompositions using {\em MathTensor}, can be outlined as
follows
\begin{itemize}
\item
Define tensors to be labeled {\em spatial}; define a vector
($n$, the unit normal) to be labeled {\em timelike} using 
functions to generate and manage lists of hypersurface-tensors,
e.g. for the ADM equations the calls would look something like:\\
{\tt
DefineSpatialVector[Shift]; DefineTimelikeVector[n, t]\\
DefineSpatial2Tensors[h, K, 1];
}

\item
Instruct {\em MathTensor} about projection rules such as 
$n^a T_{\dots a} \rightarrow 0$,
$n^a L_{n} T_{\dots a \dots} \rightarrow 0$ for $T$ spacelike.
Such rules are defined for {\em all} spatial tensors
by calling {\tt DefineHypersurfaceOrthogonalityRules[h, n]},
where $h$ is the metric induced on the hypersurface and $n$ its unit normal.

\item
The function {\tt DefineFundamentalFormsRules[h, K, a, n, Dh]}
attaches names to the geometrical objects ($n$ is the unit normal,
$Dh$ the induced covariant derivative, $a_b = n^c \nabla_c n_b$) and
defines the decomposition of the 4 metric $g_{ab} = h_{ab} +
\epsilon n_a n_b$ and the definition of the extrinsic curvature
$\nabla_a n_b = \delta \ K_{ab} + n_a a_b$ with $\epsilon$ and $\delta$
global variables to serve different sign conventions.
Many associated rules get defined automatically.

\item
Define additional rules that are particular to the problem treated.
Examples for such additional rules would be the splitting of the Maxwell tensor into 
electric and magnetic fields in Maxwell theory.
Rules for standard formulas like the Gauss-Codazzi relations or the split of the
unit normal into lapse and shift are defined by calling high-level functions.

\item Compute all independent projections of the 4-dimensional field equations
      (with respect to $h$, $n$)
      and use the projection rules defined above, afterwards
      switch to manifest three dimensional form
      (e.g. set $h_i^j = \delta_i^j$).
\item Compute equations for first order variables
      (e.g. for the Christoffel symbols).
\end{itemize}
We have defined functions to compute Lie- and covariant derivatives
for tensor densities in terms of ordinary derivatives, which is not
directly supported by {\em MathTensor}, e.g. this code defines a rule to
deal with Lie derivatives of tensor densities by adding the
appropriate correction term:
\begin{verbatim}
RuleUnique[densityLieDRule, LieD[T_[xa__], v_], (LieD[T[xa], v])
    + densityWeight[T] T[xa] OD[v[uc], lc], MemberQ[densityList, T]]
\end{verbatim}
{\tt RuleUnique} is a {\em MathTensor} command to create
rules which respect dummy indices and
{\tt densityList} is a list of all tensor density objects.

{\em Mathematica} notebooks containing examples for 3+1 decompositions, starting from
4-dimensional equations up the the generation of code for evolution
system and evaluation of constraints have been worked out for the Maxwell equations,
the ADM equations \cite{Luis}, the conformal field equations \cite{Luis}
and the BSSN equations
\cite{Luis}.
Notebooks working out the constraint propagation system have been
developed for Maxwell and the conformal field equations.
Treatment of other systems should be straightforward following these examples.

The key to generate code is to generate the lists of independent tensor components
and component equations. Sums over indices are expanded with {\em MathTensor}'s
{\tt MakeSum} command.
Assigning names to these variables as they should appear in the code
(e.g. $\{h11, h12, h13, \dots\}$) is straightforward with
Mathematica's pattern matching techniques. Ordinary derivatives
to a standard syntax which can easily be expanded to finite difference
expressions with Macros. Here we use rules such as
\begin{verbatim}
DiffCompsRule = OD[T_, n_?IntegerQ] :>
                ToExpression["D"<>ToString[Abs[n]]] @ T;
\end{verbatim}
which would yield  {\tt OD[h32, -2]} $\rightarrow$ {\tt D2h32}.
Simplifications of the resulting component expressions are obtained
with Mathematica's {\tt Collect} function.
Along these lines we developed a function to generate Cactus evolution thorns and
similar functions to generate thorns that evaluate
constraints or any user-define geometric quantities or set gauge and
initial data from a 4-metric (e.g. an exact solution).
Special care has been taken to generate nicely formatted human
readable code and to not assume a particular system of equations or
set of variables.

%\section{Summary}
%%%%%%%%%%%%%%%%%%%%%%%%%%%%%%%%%%

Summing up, we have tried to promote and discuss the use of computer algebra for
NR. CA makes it easier to focus on algorithms, detached from
a particular system of equations. Stressing a more
abstract point of view is not only mathematically more appealing but
also increases flexibility, which benefits scientific productivity.
We have very briefly outlined the content of a set of
{\em Mathematica} scripts which we developed for our use.
These scripts consist of functions to manage the bookkeeping of
hypersurface-related quantities, the definition of associated rules within
{\em MathTensor}, addition of higher-level tensor and component manipulation
functions, code generation scripts, and the development of
a set of examples to be used as templates to deal with formulations
we have not covered.
We have not produced software in the sense of user-friendly, well
documented programs with online help and extensive error-checking.
Things are not as well automated as they could be --
e.g. to deal with a new system one would essentially follow our template notebooks
instead of calling just a few high-level functions.
Some of our current code is hardwired to 3+1 decompositions, as opposed to
2+2 etc. -- it would be
interesting to generalize our code and techniques in this respect,
or even to higher dimensions.
Despite its deficiencies, we consider our work potentially useful for
others, and it is freely available on request.
%{\tt cvs -d :pserver:cvs\_anon@cvs.aei.mpg.de:/numrelcvs
%co HusaCVS/Public}
What goes much beyond our scope and resources is to build a
reasonably well documented free community tool for
computer algebra in the context of NR, e.g. comparable
to the Cactus computational toolkit -- it could however lead to a comparable
increase in productivity!
Needless to say, we have found it invaluable to accompany our computer
calculations by ``unplugged'' manual work.

{\em Acknowledgments:}
We thank the members of AEI's Astrophysical Numerical Relativity group for
stimulating discussions on the topic, S.H. has benefited from reading
{\em Mathematica} code written by Bernd Br\"ugmann and Peter H\"ubner.

\end{document}